\documentclass[12pt,preprint]{aastex}

\usepackage{epsfig}

\usepackage{xcolor}

\usepackage{amsmath,amssymb,gensymb}

\usepackage{soul}

\begin{document}

%

\title{{ Measuring Supermassive Black Hole Masses: Correlation between the Redshifts of the Fe III UV Lines and the Widths of Broad Emission Lines}}

\author{E. MEDIAVILLA\altaffilmark{1,2}, J. JIM\'ENEZ-VICENTE\altaffilmark{3,4}, J. MEJ{\'I}A-RESTREPO\altaffilmark{5}, V. MOTTA\altaffilmark{6}, E. FALCO\altaffilmark{7}, J. A. MU\~NOZ\altaffilmark{8,9}, C. FIAN\altaffilmark{1,2}  \& E. GUERRAS\altaffilmark{10}}

\altaffiltext{1}{Instituto de Astrof\'{\i}sica de Canarias, V\'{\i}a L\'actea S/N, La Laguna 38200, Tenerife, Spain}
\altaffiltext{2}{Departamento de Astrof\'{\i}sica, Universidad de la Laguna, La Laguna 38200, Tenerife, Spain}
\altaffiltext{3}{Departamento de F\'{\i}sica Te\'orica y del Cosmos, Universidad de Granada, Campus de Fuentenueva, 18071 Granada, Spain}
\altaffiltext{4}{Instituto Carlos I de F\'{\i}sica Te\'orica y Computacional, Universidad de Granada, 18071 Granada, Spain}
\altaffiltext{5}{European Southern Observatory, Alonso de C\'ordova 3107, Vitacura, Santiago, Chile.}
\altaffiltext{6}{Instituto de F\'{\i}sica y Astronom\'{\i}a, Facultad de Ciencias, Universidad de Valpara\'{\i}so, Avda. Gran Breta\~na 1111, 2360102 Valpara\'{\i}so, Chile}
\altaffiltext{7}{Harvard-Smithsonian Center for Astrophysics, 60 Garden St., Cambridge, MA 02138, USA}
\altaffiltext{8}{Departamento de Astronom\'{\i}a y Astrof\'{\i}sica, Universidad de Valencia, 46100 Burjassot, Valencia, Spain.}
\altaffiltext{9}{Observatorio Astron\'omico, Universidad de Valencia, E-46980 Paterna, Valencia, Spain}        
\altaffiltext{10}{Homer L. Dodge Department of Physics and Astronomy, The University of Oklahoma, Norman, OK, 73019, USA}

\begin{abstract}

%

We test the recently proposed {(Mediavilla et al. 2018)} black hole mass scaling relationship based on the redshift {with respect to the quasar's rest frame} of the Fe III$\lambda\lambda$2039-2113 line blend. To this end, we fit this feature in the spectra of a well suited sample of quasars, observed with X-shooter {at the Very Large Telescope (VLT)}, whose masses have been independently estimated using the virial theorem. For the quasars of this sample we {consistently} confirm  the redshift of the Fe III$\lambda\lambda$2039-2113 blend {and find that it correlates} with the squared widths of H$\beta$, H$\alpha$ and Mg II, which are commonly used {as a measure of $M_{BH}/R$} to determine masses from the virial theorem. The average differences between virial and {Fe III$\lambda\lambda$2039-2113 redshift based} masses are 0.18$\pm 0.21$ dex,  0.18$\pm 0.22$ dex and 0.14$\pm 0.21$ dex, when the {full widths at half maximum (FWHM)} of the H$\beta$, H$\alpha$ and MgII lines are, respectively, used. The difference is reduced to 0.10$\pm 0.16$ dex when the standard deviation, $\sigma$, of {the} MgII {line} is used, instead.  {{We also study} the high S/N composite quasar spectra of the {Baryon Oscillation Spectroscopic Survey (BOSS)}, finding that the Fe III$\lambda\lambda$2039-2113 redshifts and Mg II {squared widths}, $FWHM_{MgII}^2$, match very well the correlation found for the individual quasar spectra observed with X-shooter}. {This correlation is expected} if the redshift is gravitational. 


\end{abstract}


\section{Introduction \label{intro}}

The masses of {Super-Massive Black Holes (SMBH)} found in the center of galaxies are known to correlate strongly with the properties of their host galaxies (Kormendy \& Richstone 1995, Magorrian et al. 1998). Understanding the coupling between the growth of {SMBH} and the evolution of galaxies requires measuring {SMBH} masses at early epochs, i.e., in distant quasars (Peng et al. 2006).
In quasars, a central {SMBH} is surrounded by a disc of inspiralling matter (Zeldovich 1964, Salpeter 1964)  that  illuminates gas clouds located in a larger region (Broad Line Region, BLR), giving rise to broad emission lines (BEL).  In this scenario, the {SMBH} mass can be estimated assuming virialization of the BLR gas clouds (Peterson 2014). However, the accuracy of individual virial mass estimates is limited to $\sim0.4\rm\,dex$ (e.g. Peterson 2014) by our ignorance about the BLR structure and dynamics. To complicate matters further, in the case of distant quasars, the emission line of reference observable in the optical is CIV, whose suitability to estimate masses is controversial (see, e.g., Mej{\'{\i}}a-Res{trepo et al. 2016 and references therein).

Another approach to measure masses is the gravitational redshift that the {SMBH} should induce on the BEL (Netzer 1977, Anderson 1981, Mediavilla \& Moreno-Insertis 1989). If the kinematics is ruled by gravitation, the typical widths of the  BEL, {$\Delta\,v\lesssim5\times 10^3\rm\,km\,s^{-1}$, imply gravitational redshifts, $\Delta\,z_{grav}\lesssim0.0004$, which correspond to wavelength redshifts of $\lesssim1\,\rm \AA$ for the UV emission lines, and of $\lesssim 2\rm\, \AA$ for the optical ones. However, owing to the complex morphology of the BEL, which are typically blends of several  components with different kinematics and origin, the detection of shifts of this magnitude ({perhaps affecting only one of the components}) is quite difficult. Likely for this reason}, detections of gravitational redshift using the typical strong BEL (Peterson et al. 1985, Sulentic 1989, Zheng \& Sulentic 1990, Popovic et  al. 1995, Corbin 1997, Kollatschny 2003, Jonic et al. 2016, Liu et al. 2017) have been, until now, scarce and inaccurate. {The situation is particularly bad in the case of distant quasars for which the available data at (rest frame) optical wavelengths are limited}.

{In this context the recent report by Mediavilla et al. (2018) of a {consistently observed} redshift\footnote{Here and hereafter,  we refer to the Fe III$\lambda\lambda$2039-2113 redshift with respect to the quasar's rest frame, i.e., after correcting for the cosmological redshift.} of the Fe III$\lambda\lambda$2039-2113 {blend\footnote{{This feature} is weak but, according to Vestergaard \& Wilkes 2001, relatively free of contamination from lines of other species.}} by amounts explicable by the gravitational redshift induced by the {SMBH},  is promising.    Based on the impact of microlensing magnification, {this feature} seems to arise from an inner region of the BLR of about 13 light-days {in} size (Fian et al. 2018) for quasars of average luminosity $\langle \lambda L_\lambda(1350\rm \AA)\rangle=10^{45.8}\,erg\,s^{-1}${. This size, smaller by more than one order of magnitude than the typical dimensions of the BLR, would explain the relatively large measured redshifts}. From these results, in Mediavilla et al. (2018) we calibrate a mass scaling relationship based on the Fe III$\lambda\lambda$2039-2113 redshift useful to measure the {SMBH} mass from a single spectrum\footnote{It is important to remark that this alternative way to estimate masses is not only simple to apply but also free from geometrical effects and largely insensitive to nongravitational forces.}. The estimated masses are in good agreement with virial masses over two orders of magnitude.

The objective of this {work is} to test {the use of the Fe III$\lambda\lambda$2039-2113 redshift to estimate {SMBH} masses, studying} a sample of quasars with independent virial mass estimates. The objects in this sample should have a convenient redshift (to exhibit the Fe III UV lines in the optical), S/N ratio as high as possible (the Fe III UV blend is relatively weak) and measured virial mass. The sample of 39 unobscured quasars at $z\sim 1.5$ observed with X-shooter@VLT by Capellupo et al. (2015, 2016), selected to cover a wide range in masses and Eddington ratios is, to our knowledge, the best available dataset for this purpose. At the redshift of the sample, the broad X-shooter wavelength range allows observations of the four strong lines (C IV, Mg II, H$\alpha$ and H$\beta$) commonly used to estimate virial masses, and also includes the Fe III$\lambda\lambda$2039-2113 blend. Virial masses of the black holes of the quasars in the sample have been derived by Mej{\'{\i}}a-Restrepo et al. (2016).

The paper is organized as follows: In \S 2 we describe the fits of the Fe III$\lambda\lambda$2039-2113 blend in the subsample of objects (10) with sufficient S/N ratio. In this section we also estimate the Fe III$\lambda\lambda$2039-2113  line shifts {{with respect to the quasars'} rest frame} and study their dependence {on} the H$\beta$, H$\alpha$ and Mg II emission line {widths}. In \S 3 we derive the masses from the {Fe III$\lambda\lambda$2039-2113}{ redshifts and compare {them} with the virial masses. In this section we also infer a size-luminosity relationship for the Fe III$\lambda\lambda$2039-2113 emitting region. Finally, in \S 4 we summarize our main conclusions.

\section{Results}

\subsection{Fe III$\lambda\lambda$2039-2113 Redshift Measurements \label{proxy1}} 

The S/N of the spectra studied by Mej{\'{\i}}a-Restrepo et al. (2016) is high (S/N $>$100) in the region of the Fe III$\lambda\lambda$2039-2113 blend. However, after continuum subtraction not all the spectra present a Fe III$\lambda\lambda$2039-2113 feature with S/N  high enough to allow {for} a reasonably good fit. We have selected the 10 systems (J0019-1053, J0043+0114,  J0155-1023, J0209-0947, J0404-0446, J0842+0151, J0934+0005, J0941+0443, J1002+0331 and J1158-0322) with 
$S/N>5$ at the peak of the blend after continuum subtraction, and without features that prevent the determination of the continuum around the Fe III$\lambda\lambda$2039-2113 blend. We model the blend following the same steps as in Mediavilla et al. (2018), although, when convenient, the windows used to fit the continuum have been adapted to the shape of the continuum, which can change from object to object\footnote{The fits match well the shape of the blend with reduced chi-squared values $\chi^2_{red}<1.4$.}. We have used the peak of the Mg II emission line to {define} the quasar cosmological redshift. We have checked the consistence of this determination with the narrow H$\beta$ (in all the spectra) and [OIII] (in four objects) emission lines finding differences from the rest wavelengths smaller than the errors in the fit ($\sim1$\AA). 

After correcting the spectra {for the cosmological redshifts}, we find that the Fe III$\lambda\lambda$2039-2113 blend is redshifted in the 10 quasars of the sample. This confirms the {consistently observed} redshift of this feature in quasars found by Mediavilla et al. (2018).  We estimate an average redshift and scatter among the 10 spectra of $\langle \Delta \lambda \rangle=5.5\pm 2.9\rm\, \AA$, comparable to the redshift of $\sim 7\rm\, \AA$ found in the high S/N composite SDSS spectrum (Mediavilla et al. 2018). 

\subsection{Correlation between the Fe III$\lambda\lambda$2039-2113 Redshifts and  the H$\beta$, H$\alpha$ and Mg II Squared Widths\label{proxy2}}{

In the upper left panel of Figure \ref{grid1} we plot the squared widths of H$\beta$, $FWHM_{H\beta}^2$, measured by  Mej{\'{\i}}a-Restrepo et al. (2016) vs. the Fe III UV wavelength redshifts, $\Delta \lambda$, for the 10 objects in the sample. There is a significant correlation between both quantities (linear fit with Pearson correlation coefficient, $r_s=0.61$), thus supporting that the redshift is related to $M_{BH}/R$, as is the case of the squared widths. Considering that $FWHM_{H\beta}^2$ is the preferred quantity to infer masses using the virial theorem, this correlation {supports} the gravitational origin of the measured redshifts.

{In principle, alternative explanations leaving aside the gravitational redshift scenario are possible.  Kova{\v c}evi{\'c}-Doj{\v c}inovi{\'c} \& Popovi{\'c} (2015) interpret the redshifts found in the case of the UV Fe II emission lines as inflow of gas clouds located at the outer parts of the BLR. According to Ferland et al. (2009), infall requires that the clouds have large column densities and, hence, we will see predominantly the shielded face of the near-side infalling clouds with the consequent redshift of the emission lines. However, this or other alternative hypothesis should provide a convincing explanation of the good correlation between the observed redshifts and  $FWHM_{H\beta}^2$.}

The correlation between widths and redshifts is also observed in H$\alpha$ and in Mg II (Fig. \ref{grid1}) with correlation coefficients  $r_s=0.69$ and $r_s=0.62$, respectively. It is interesting to note that, in the case of Mg II, the correlation becomes very strong ($r_s=0.82$) when we take the squared standard deviation calculated by Mej{\'{\i}}a-Restrepo et al. (2016), $\bf \sigma^2$,  instead of $FWHM^2$ as the magnitude related to the virial {masses} (Fig. \ref{grid1}). On the contrary, in the case of H$\beta$, the correlation is weak if we adopt $\sigma^2$ as the virial indicator. {In the case of Mg II, the improvement may be explained by the higher precision measure of the virial product obtained by using {$\sigma^2$ (instead of $FWHM^2$)} (Peterson et al. 2004), or by the larger bias (likely related to inclination) of $FWHM^2$ based measurements of SMBH masses with respect to those based on $\sigma^2$ (Collin et al. 2006). In the case of the more complex H$\beta$ line profiles, these factors could be less significant than the presence of} several components at low levels of intensity, which can be affecting in a very different way $\sigma$ and FWHM. In fact, the correlation between $\sigma$ and FWHM is by far worst in the case of H$\beta$.

%
%
%

{
\subsection{Correlation between the Fe III$\lambda\lambda$2039-2113 Redshift and  $FWHM_{MgII}^2$ for BOSS Composite Spectra\label{proxy3}}
Encouraged by the correlations found  in the previous section between the Fe III$\lambda\lambda$2039-2113 redshifts and the squared widths of several emission lines, we check these results with the high S/N composite spectra from the BOSS survey (Jensen et al. 2016). Mediavilla et al. (2018) obtained the Fe III$\lambda\lambda$2039-2113 redshifts for the BOSS composite spectra with high enough S/N ratio. For 13 of these spectra, the wavelength coverage of {the} BOSS survey includes the MgII line (H$\beta$ and H$\alpha$ are out of the observed wavelength range). 

We estimate the $FWHM_{MgII}$ in a direct way, fitting the continuum to a straight line in the $\lambda\lambda$2650-2700 and $\lambda\lambda$2900-2925  wavelength ranges and measuring the width at 50\% of the peak intensity {with} respect to this continuum. The resulting FWHMs have a strong correlation ($r_s=0.77$) with the Fe III$\lambda\lambda$2039-2113 redshifts (upper panel in Figure \ref{figcomp}). Very interestingly, the BOSS data match very well the correlation obtained with the data from Mej{\'{\i}}a-Restrepo et al. (2016) in  \S \ref{proxy2} (see lower panel in Figure \ref{figcomp}). {Note that in Figure 2 we have divided the FWHM of BOSS data by a factor 1.06 to equal the means of the squared widths, $\langle FWHM^2\rangle$, of both data sets\footnote{{This slight global shift does not affect the slope of the fit to the BOSS data and would correspond to a shift by a factor of 1.12 in mass estimates, which is also small given the typical uncertainties in virial masses.}}}. Notice  the excellent agreement between the linear fits to the BOSS, Mej{\'{\i}}a-Restrepo et al. (2016) and all data.


The good correlation of the Fe III$\lambda\lambda$2039-2113 redshifts with the $FWHM_{MgII}^2$ of the BOSS composites and the match with the correlation obtained in \S \ref{proxy2} from Mej{\'{\i}}a-Restrepo et al. (2016) data, confirm the hypothesis that the Fe III$\lambda\lambda$2039-2113 redshifts are a proxy for {$M_{BH}/R$}.
}
{\section{Discussion}

\subsection{Test of the {SMBH} Mass Scaling Relationship Based on the Fe III$\lambda\lambda$2039-2113 Redshift \label{RM}}

To test the {SMBH} mass measurements from the Fe III$\lambda\lambda$2039-2113 redshift, we consider the mass scaling relationship of Mediavilla et al. (2018) based on the best fit to the pairs, $(M_{vir},M_{{redshift}})$ of a sample of 10 AGN and quasars, using as free parameter the exponent of the R-L relationship, $R\propto \lambda L_\lambda^b$, (Eq. 14 of Mediavilla et al. 2018):

\begin{equation}
\label{nonlinear}
M^{Fe III}_{BH}=10^{7.89^{+0.11}_{-0.13}}\left(z_{Fe III}c\over 10^3\,{\rm km\,s^{-1}}\right)\left({\lambda L_\lambda (1350{\rm\,\AA})\over 10^{44} {\rm\, erg\,s^{-1}}}\right)^{0.57\pm0.08}M_\odot.
\end{equation}
We use this equation to estimate the masses {based on the Fe III$\lambda\lambda$2039-2113 redshift} of the 10 objects in the sample of Mej{\'{\i}}a-Restrepo et al. (2016)  and compare them with the virial masses obtained by Mej{\'{\i}}a-Restrepo et al. (2016) from the H$\beta$, H$\alpha$  and Mg II {FWHMs}. The results are shown in Figure \ref{grid2}. The agreement is rather good, with mean shifts {and scatters ($\pm1\,\sigma$)} between the {redshift based} and virial masses of $0.18\pm 0.21$, $0.18\pm0.22$ and $0.14\pm 0.21$ dex for H$\beta$, H$\alpha$ and Mg II, respectively. If we take $\sigma$ as reference for the Mg II widths instead of the FWHM, we obtain a significantly better fit (see Figure \ref{grid2}) with a mean shift of 0.10 dex and  a scatter of $\pm 0.16$ dex. Notice that these scatters are comparable to {the intrinsic scatter between virial masses obtained from} H$\alpha$ {\bf($\pm\, 0.16$)} and Mg II {\bf($\pm\, 0.25$)} with respect to H$\beta$ (Mej{\'{\i}}a-Restrepo et al. 2016). {In other words, for the studied sample, the intrinsic uncertainty of virial masses is an upper limit of the uncertainty of mass estimates based on the redshift.}

Although the mean differences between the  {redshift} and virial {based} masses are below the scatter, on average the  {redshift} based masses are smaller than the virial based ones.  This may come from the derivation of  Eq. \ref{nonlinear}. The virial masses used by Mediavilla et al. (2018) to calibrate this Fe III UV mass scaling relationship come from different bibliographic sources and are based on a variety of emission lines. The use of different emission lines or different procedures to evaluate the line widths introduce systematics that could explain the offset. In fact, in our case the offset is significantly smaller when $\sigma_{MgII}^2$ is used. Leaving aside the statistical significance of the offsets, it is important to mention that some recent results indicate that mass determinations based on reverberation mapping are, on average, smaller than those based on single epoch measurements and the R-L relationship (Grier et al. 2017, Lira et al. 2018). 

After the consistent results described above,  we may recalibrate the mass scaling relationship with Fe III$\lambda\lambda$2039-2113 redshift (Eq. \ref {nonlinear}) using the 10 objects of the sample {in Mediavilla et al. (2018) plus the new 10 objects from Mej{\'{\i}}a-Restrepo et al. (2016). The change would imply a{ relatively small offset in the calibration of $\sim 0.09$ dex in the case of H$\beta$ and of only $\sim 0.05$ dex for $\sigma_{MgII}$. In any case, any recalibration of Eq. \ref{nonlinear} basically implies a recalibration of the size-luminosity, R-L, relationship for Fe III$\lambda\lambda$2039-2113. We address this issue in the next section.

\subsection{Size-Luminosity Relationship for Fe III$\lambda\lambda$2039-2113\label{R-L}}

The correlation between $FWHM_{H\beta}^2$ and the redshift of the Fe III$\lambda\lambda$2039-2113 blend also implies a correlation between the sizes of the regions emitting H$\beta$, $R_{H\beta}$, and the Fe III UV blend, $R_{FeIII}$.  Combining the virial equation for H$\beta$,

\begin{equation}
\label{virial}
M_{BH}\simeq f_{H\beta}{FWHM_{H\beta}^2R_{H\beta}\over G},
\end{equation}
where $f_{H\beta}$ is the virial factor (which may change from object to object), with the equation relating the {SMBH} mass with the gravitational redshift of the Fe III UV lines (see Eq. 3 in Mediavilla et al. 2018),

\begin{equation}
\label{mass}
M_{BH}\simeq{2 c^2\over3 G}{\Delta \lambda\over \lambda}{R_{FeIII}},
\end{equation}
we obtain,

\begin{equation}
\label{linear}
{R_{FeIII}\over R_{H\beta}}\simeq{3 \over 2}f_{H\beta}\left(FWHM_{H\beta}/c\right)^2 \left({\Delta \lambda\over \lambda}\right)^{-1}.
\end{equation}
An immediate result of this equation is that the observed correlation between $FWHM_{H\beta}^2$ and ${\Delta \lambda}$ makes sense only if the  ${1\over f_{H\beta}}{R_{FeIII}\over R_{H\beta}}$ factor does not change very much from object to object. {As there is no reason to expect that individual variations in  $f_{H\beta}$ are compensated by simultaneous variations in ${R_{FeIII}\over R_{H\beta}}$ that leave the whole factor unchanged, we conclude that both factors must be approximately constant among the different objects.} Averaging Eq. \ref{linear} over the 10 objects selected from  the Mej{\'{\i}}a-Restrepo et al. (2016) sample  we obtain:

\begin{equation}
\label{linear2}
\Bigl<{ {R_{FeIII}\over R_{H\beta}}\Bigr>}\simeq{3 \over 2}f_{H\beta}\Bigl< {\left(FWHM_{H\beta}/c\right)^2 \left({\Delta \lambda\over \lambda}\right)^{-1}} \Bigr>=(0.17\pm0.08) f_{H\beta},
\end{equation}
(Notice that this value corresponds to a mean luminosity of the quasars of $\langle \lambda L_\lambda(1350\rm \AA)\rangle=10^{46.35}\,erg\,s^{-1}$). Then, for the value $f_{H\beta}=1$ adopted by Mej{\'{\i}}a-Restrepo et al. (2016), the size of the Fe III UV emitting region is significantly smaller than the size associated with H$\beta$. {Similar scaling relationships can be obtained for H$\alpha$, $\langle{ {R_{FeIII}/R_{H\alpha}}\rangle}\simeq (0.15\pm0.07) f_{H\alpha}$, and MgII, $\langle{ {R_{FeIII}/R_{MgII}\rangle}\simeq (0.08\pm0.04) f_{MgII}}$.}

According to Mej{\'{\i}}a-Restrepo et al. (2016) the scaling with luminosity of the H$\beta$ emitting region is given by  $R_{H\beta}\propto (\lambda L_\lambda)^{0.650}$.  {Taking into account this dependence in Equation \ref{mass}  and} using the {SMBH} mass {averaged over the 10 objects selected from the} Mej{\'{\i}}a-Restrepo et al. (2016) {sample corresponding to} H$\beta$ we can obtain a scaling relationship of size with luminosity for Fe III UV,

\begin{equation}
\label{mass2}
{R_{FeIII}}\simeq \left({\lambda L_\lambda(1350\rm \AA)\over 10^{45.79}\,erg\,s^{-1}}\right)^{0.65} 27.1^{+9.5}_{-3.2}\,\rm light-days
\end{equation}
where, according to the $R_{FeIII} \propto R_{H\beta}$ proportionality, we have used the 0.65 exponent inferred by  Mej{\'{\i}}a-Restrepo et al. (2016). Mediavilla et al. (2018) reached a similar result. {Writing Equation 13 from Mediavilla et al. (2018) in the same way as Equation 11 of these authors and identifying terms, we obtain,}
\begin{equation}
\label{mass3}
{R_{FeIII}}\simeq \left({\lambda L_\lambda(1350\rm \AA)\over 10^{45.79}\,erg\,s^{-1}}\right)^{0.57} 19.6^{+8.3}_{-2.2}\,\rm light-days.
\end{equation}
In principle, we could use the R-L scaling relationship of Eq. \ref{mass2} to recalibrate Eq. \ref{nonlinear} to remove the (relatively small) offset between the Mej{\'{\i}}a-Restrepo et al. (2016) virial masses and our {Fe III$\lambda\lambda$2039-2113 redshift based} estimates. However, recent size estimates based on reverberation mapping (Grier et al. 2017) are on average smaller than the single epoch ones based on the R-L relationship, especially for high luminosity objects like the ones in the Mej{\'{\i}}a-Restrepo et al. (2016) sample. Then, it seems preferable to maintain the calibration of Eq. \ref{nonlinear} until this discrepancy is resolved. 

{Notice that Equations \ref{mass} to \ref{mass2}  can also be seen as a consequence of the correlation found between redshifts, ${\Delta \lambda/ \lambda}$,  and squared widths, $FWHM^2$, without the need of invoking the gravitational redshift hypothesis although, then, the $2/3$ factor would be different. On the other hand, Eq. \ref{mass3} is an empirical result independent of the origin of the redshifts.

The question of the localization of the Fe III$\lambda\lambda$2039-2113 emission region is crucial to decide between several possible hypothesis about the origin of the redshifts. Reverberation mapping is, indeed, the most robust and independent way to measure the size of the emission region. Until the results of this observational technique become available,  the outcomes of the present work can be regarded like empirical findings,  irrespective of the origin of the measured redshifts.
}

\section{Conclusions}

We fit the Fe III$\lambda\lambda$2039-2113 emission line blend in a sample of 10 quasars with the best spectra available to study this feature (Mej{\'{\i}}a-Restrepo et al. 2016). {We have extended part of this analysis to the composite quasar spectra from the BOSS survey}. The results support {that the} redshift of these lines {is a proxy for $M_{BH}/R$ and} the possibility of using this redshift to estimate {SMBH} masses. Specifically:

1 - We  {consistently} confirm the redshift of the Fe III$\lambda\lambda$2039-2113 feature. We find an average redshift of $\langle \Delta \lambda \rangle \simeq 5.5$\AA, well above uncertainties ($0.9$\AA).

2 - There is a linear correlation ($r_s\gtrsim 0.6$) between the measured Fe III$\lambda\lambda$2039-2113 redshift and the squared {widths} ($FWHM^2$) of the H$\beta$, H$\alpha$ and Mg II lines independently determined by Mej{\'{\i}}a-Restrepo et al. (2016). This correlation{, nicely confirmed in the case of Mg II by the analysis of the  high S/N ratio composite quasar spectra from the BOSS survey ($r_s=0.77$),}  supports the hypothesis that {these redshifts measure $M_{BH}/R$}{. The correlation with the squared widths is expected if} the origin of the Fe III$\lambda\lambda$2039-2113 redshift is gravitational. The correlation is surprisingly tight ($r_s=0.82$) with $\sigma_{Mg II}^2$.

3 - The scaling relationship of mass with redshift and luminosity given by Mediavilla et al. (2018) predicts the masses for the {SMBH} associated with the quasars of the sample of  Mej{\'{\i}}a-Restrepo et al. (2016) with mean offsets of 0.18$\pm 0.21$ dex  respect to the virial masses computed from $FWHM_{H\beta}$,  0.18$\pm 0.22$ dex (virial masses from $FWHM_{H\alpha}$), 0.14$\pm 0.21$ dex (virial masses from $FWHM_{MgII}$) and 0.10$\pm 0.16$ dex (virial masses from $\sigma_{MgII}$). The statistical agreement between  {redshift based} and virial masses is as good as the internal agreement among virial mass estimates based on different lines (and/or definitions of the line widths).

4 - The correlation between widths and  {Fe III$\lambda\lambda$2039-2113} redshifts provides a size-luminosity, R-L, scaling for $R_{Fe III}$ in reasonable agreement with the results from Mediavilla et al. (2018).

The calibration of the mass scaling relationship with the  {Fe III$\lambda\lambda$2039-2113} redshift resides on the size of the region emitting the Fe III$\lambda\lambda$2039-2113. Until now, this size has been derived (Mediavilla et al. 2018)  either directly from microlensing (an average estimate with large uncertainties) or, indirectly, matching the {redshift based} and virial masses (obtained using emission lines arising from very much larger regions). The measurements of a sample of quasars made in the present work support these results. However, reverberation mapping observations of  Fe III$\lambda\lambda$2039-2113 are needed to determine, in an independent and accurate way, the size of the emitting region and, if possible, to obtain a size-luminosity, R-L, relationship for this blend.
}
\acknowledgements{We thank the anonymous referee for the thorough review, valuable comments and suggestions. We thank the SDSS and BOSS surveys for kindly providing the data. This research was supported by the Spanish MINECO with the grants AYA2016-79104-C3-1-P and AYA2016-79104-C3-3-P.  J.J.V. is supported by the projects AYA2014-53506-P and AYA2017-84897-P financed by the Spanish Ministerio de Econom\'\i a y Competividad and by the Fondo Europeo de Desarrollo Regional (FEDER), and by project FQM-108 financed by Junta de Andaluc\'\i a.  V.M. gratefully acknowledges support from  Centro de Astrof\'\i sica de Valpara\'\i so. C.F. acknowledges the financial support of La Caixa fellowship.}

%
%
%

\clearpage

\begin{figure}[h]
\includegraphics[scale=0.85]{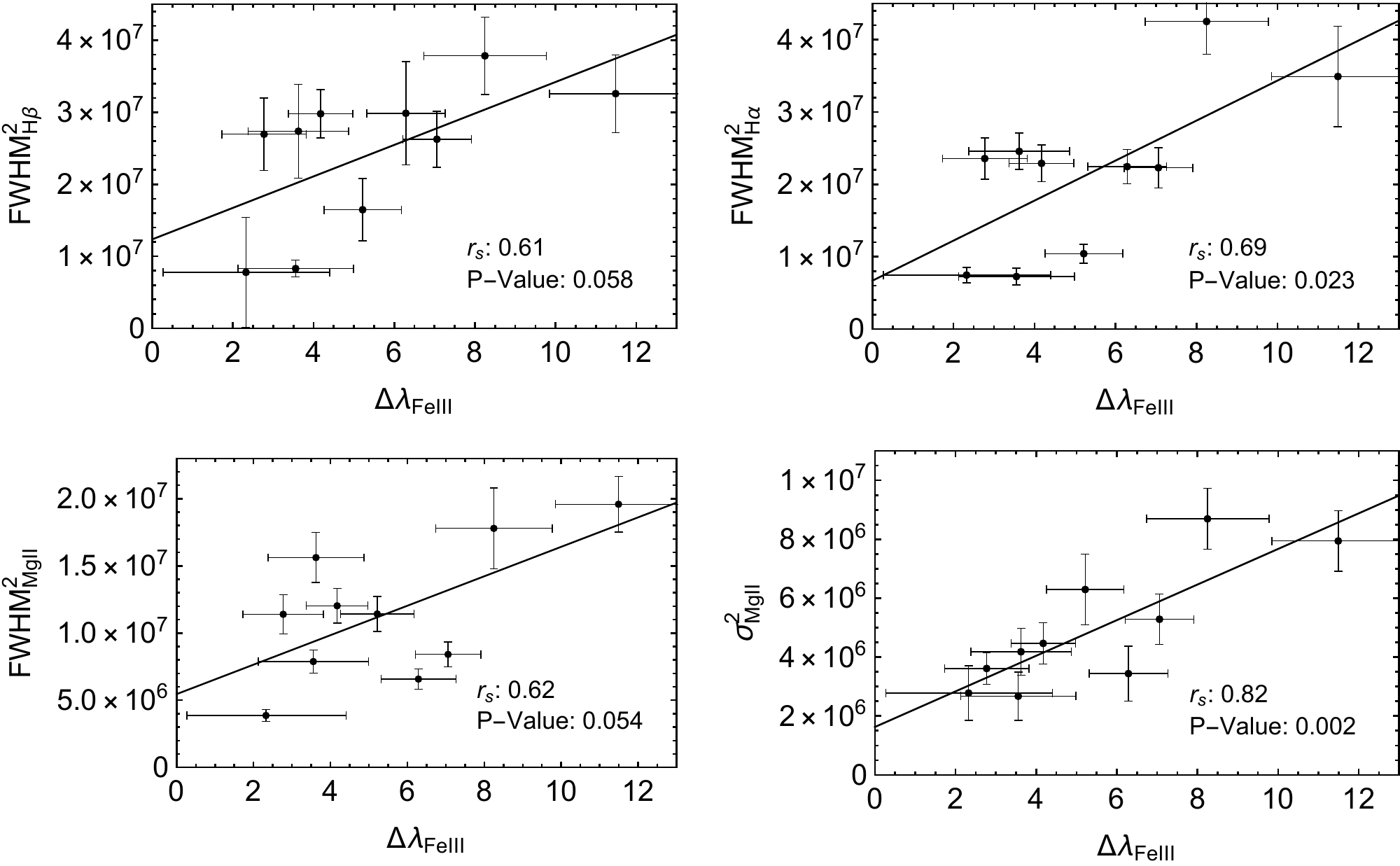}
\caption{Squared widths of several emission lines (from top to bottom, left to right, $FWHM_{H\beta}^2$ , $FWHM_{H\alpha}^2$,  $FWHM_{MgII}^2$ and $\sigma_{MgII}^2$) versus wavelength redshift, ${\Delta \lambda}$, of the UV iron blend (Fe III$\lambda\lambda$2039-2113) for Mej{\'{\i}}a-Restrepo et al. (2016) quasars. Straight lines correspond, in each case, to the best linear fit 
(see text). \label{grid1}}
\end{figure}

\begin{figure}[h]
\includegraphics[scale=0.95]{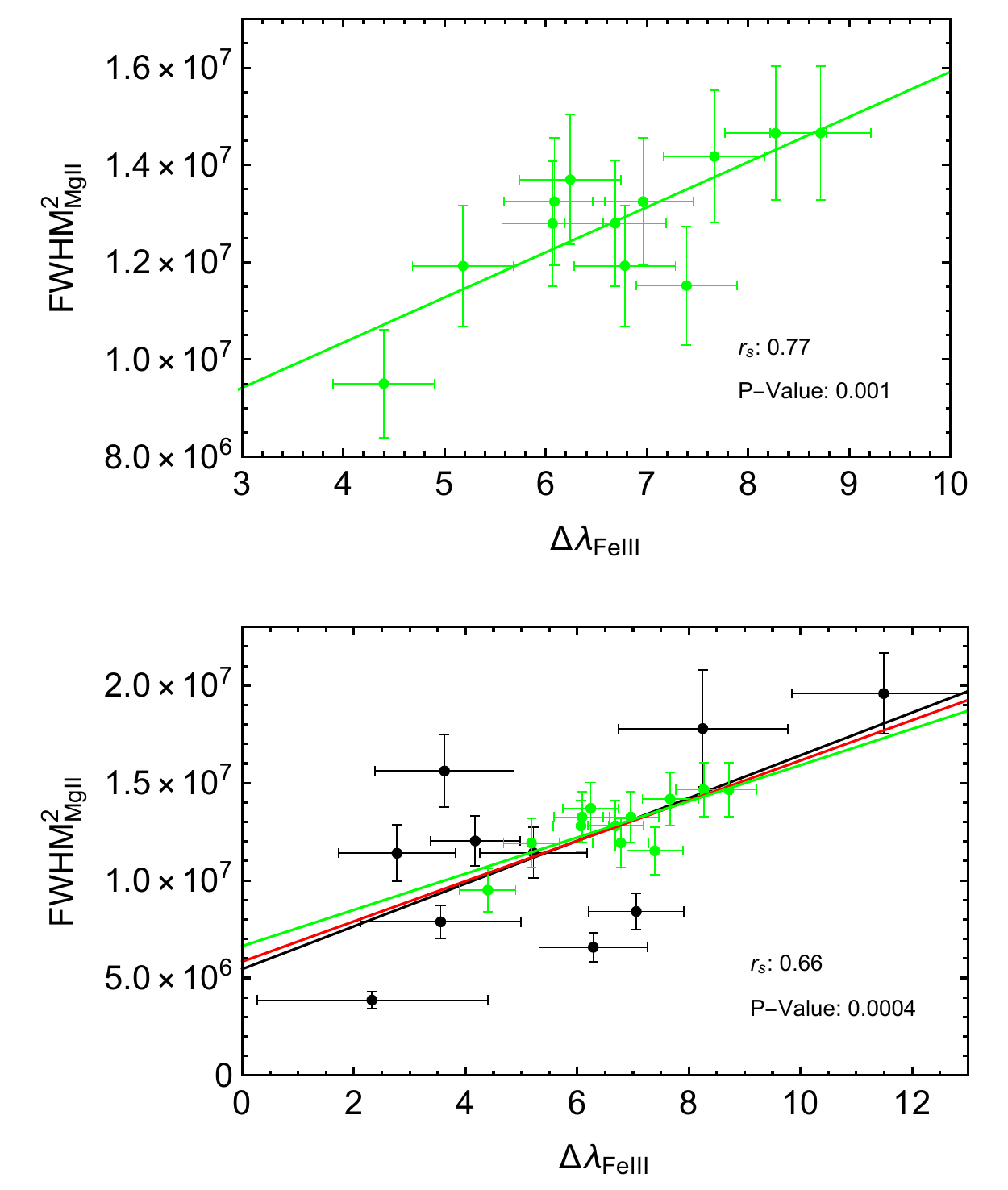}
\caption{{$FWHM_{MgII}^2$ versus wavelength redshift, ${\Delta \lambda}$, of the UV iron blend (Fe III$\lambda\lambda$2039-2113). The green and black points correspond to BOSS quasar composites (Mediavilla et al. 2018) and to Mej{\'{\i}}a-Restrepo et al. (2016) quasars, respectively. Upper panel: $FWHM_{MgII}^2$ of BOSS composites; the green line is the best linear fit to the data {(notice that we have divided the FWHM of BOSS data by a factor 1.06 to match the means of the squared widths, $\langle FWHM^2\rangle$, of BOSS and Mej{\'{\i}}a-Restrepo et al. data sets)}.  Lower left panel: comparison between Mej{\'{\i}}a-Restrepo et al. (2016) and BOSS composites data; the green, black and red straight lines are the best linear fits to the BOSS, Mej{\'{\i}}a-Restrepo et al. (2016), and all data, respectively
(see text). P-value and $r_s$ in the lower panel correspond to the fit to all the data.} \label{figcomp}}
\end{figure}

\begin{figure}[h]
\includegraphics[scale=0.85]{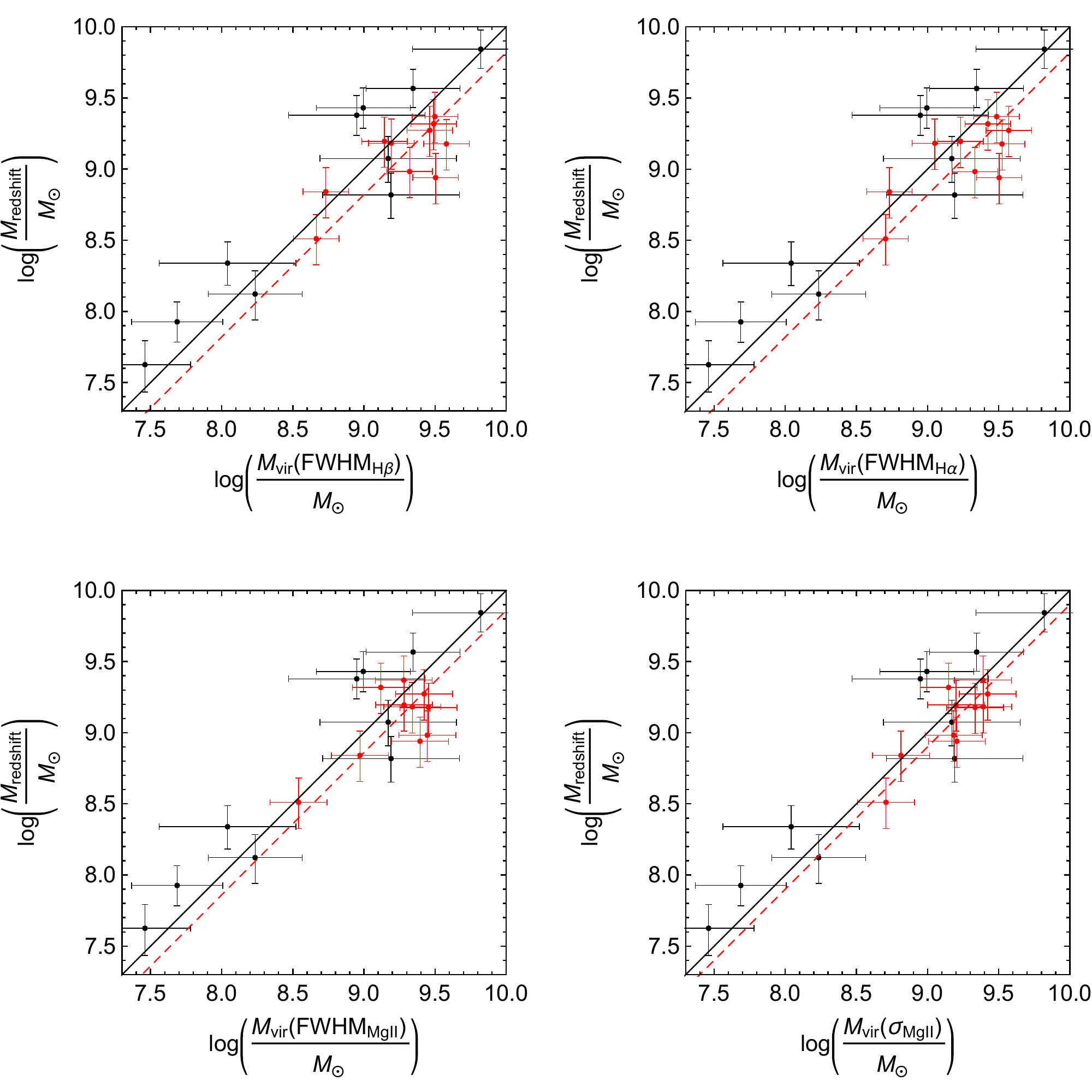}
\caption{Comparison between the virial masses (from top to bottom, left to right: $M_{vir}(FWHM_{H\beta})$, $M_{vir}(FWHM_{H\alpha})$, $M_{vir}(FWHM_{MgII})$ and $M_{vir}(\sigma_{MgII})$ and {Fe III$\lambda\lambda$2039-2113} redshift based masses. In black the original data used in the calibration of the redshift mass scaling (Mediavilla et al. 2018). In red the new data from Mej{\'{\i}}a-Restrepo et al. (2016). Solid lines correspond to $M_{redshift}=M_{vir}$. Dashed lines correspond, in each case, to the best linear fit with slope unity to Mej{\'{\i}}a-Restrepo et al. (2016) data. Errors in $M_{vir}$ for the new data correspond to the scatter of the virial relationships. Errors in $M_{redshift}$ for the new data include the errors in the parameters of the fit and 0.13 dex of intrinsic  scatter in the R-L relationship (Peterson 2014). \label{grid2}}
\end{figure}

\end{document}